\documentclass[journal=jacsat,manuscript=article]{achemso}

\usepackage{chemformula}
\usepackage[T1]{fontenc}

\usepackage{graphicx}
\usepackage{amsmath, amsfonts, amssymb, bm}
\usepackage{color}
\usepackage{multirow}
\usepackage[utf8]{inputenc}
\usepackage[spanish,USenglish]{babel}
\usepackage{amsmath,array,amssymb,pst-all}
\usepackage{dsfont} 
\usepackage{bbold}     
\usepackage{ragged2e}

\setkeys{acs}{articletitle = true}

\author{Duvalier Madrid-Úsuga}
\affiliation[Universidad del Valle]
{Centre for Bioinformatics and Photonics---CIBioFi, Calle 13 No. 100-00, Edificio E20, No. 1069, Universidad del Valle, Cali 760032, Colombia}
\alsoaffiliation[Universidad del Valle]
{Department of Physics, Universidad del Valle, 760032 Cali, Colombia}
\email{duvalier.madrid@correounivalle.edu.co}
\author{Carlos A. Melo-Luna}
\affiliation[Universidad del Valle]
{Centre for Bioinformatics and Photonics---CIBioFi, Calle 13 No. 100-00, Edificio E20, No. 1069, Universidad del Valle, Cali 760032, Colombia}
\alsoaffiliation[Universidad del Valle]
{Department of Physics, Universidad del Valle, 760032 Cali, Colombia}
\author{Alberto Insuasty}
\affiliation[Universidad del Norte]
{Department of Chemistry and Biology, Universidad del Norte, Km 5 via Puerto Colombia, 081007 Barranquilla, Colombia}
\author{Alejandro Ortiz}
\affiliation[Universidad del Valle]
{Centre for Bioinformatics and Photonics---CIBioFi, Calle 13 No. 100-00, Edificio E20, No. 1069, Universidad del Valle, Cali 760032, Colombia}
\alsoaffiliation[Universidad del Valle]
{Department of Chemistry, Universidad del Valle, 760032 Cali, Colombia}
\author{John H. Reina}
\affiliation[Universidad del Valle]
{Centre for Bioinformatics and Photonics---CIBioFi, Calle 13 No. 100-00, Edificio E20, No. 1069, Universidad del Valle, Cali 760032, Colombia}
\alsoaffiliation[Universidad del Valle]
{Department of Physics, Universidad del Valle, 760032 Cali, Colombia}
\email{john.reina@correounivalle.edu.co}
\title[An \textsf{achemso} demo]
  {Optical and Electronic Properties of Molecular Systems Derived from Rhodanine}
\abbreviations{IR,NMR,UV}
\keywords{American Chemical Society, \LaTeX}

\begin{document}
\begin{tocentry}
\vspace{-0.4 cm}
\begin{figure}[H]
\begin{center}
\includegraphics[height=4.0cm, width=6.5cm]{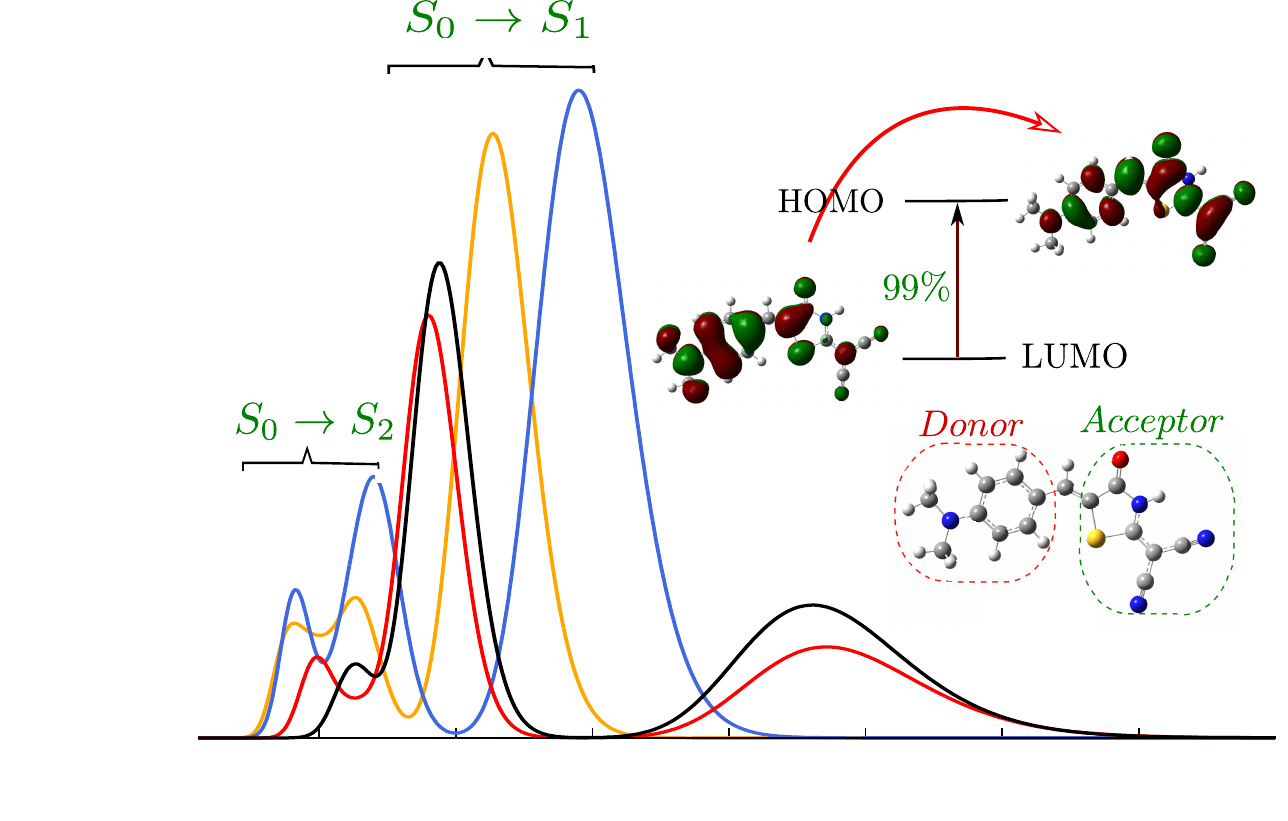}
\end{center}
\end{figure}
\end{tocentry}

\begin{abstract}
  Push-Pull functional compounds consisting of dicyanorhodanine derivatives have attracted a lot of interest because their optical, electronic, and charge transport properties make them useful as building blocks for organic photovoltaic implementations. The analysis of the frontier molecular orbitals shows that the vertical transitions of electronic absorption are characterized as intramolecular charge transfer; furthermore, we show that the analyzed compounds exhibit bathochromic displacements when comparing the presence (or absence) of solvent as an interacting medium. In comparison with materials defined by their energy of reorganization of electrons (holes) as electron (hole) transporters, we find a transport hierarchy whereby the molecule (Z)-2-((1,1-Dicyanomethylene)-5-(4-dimethylamino)benzylidene)-1,3-thiazol-4 is better at transporting holes than molecule (Z)-2-((1,1-Dicyanomethylene)-5-(tetrathiafulvalene-2-ylidene)-1,3-thiazol-4.
  
\end{abstract}

\section{Introduction}
Charge transfer (CT) studies seek to understand the ways in which their transfer rate of CT depends on the properties of the electron-donor and electron-acceptor system, solvent, molecular bridge and electronic coupling between the involved states \cite{genereux2009, delor2015}. The different functionality played by these factors and the way they affect the qualitative and quantitative aspects of the electron transfer process have been extensively discussed in recent years\cite{Nitzan, Zamadar, Renaud, Guan, Belmonte}. The need for understanding the processes of electron (ET) or charge transfer at the molecular level have prompted the study of highly conjugated molecular systems of the donor-acceptor (D-A) type, given their unique photo-physical and photo-activated properties {\cite{lee2011new}}. 

These properties have favoured the application and development of such systems in areas that comprise non-linear optical materials {\cite{dragonetti2016}}, molecular optical switches{\cite{Kanaani2018}}, and photovoltaic cells{\cite{kesters2015}}, among others. In the field of photovoltaics, organic photovoltaic devices with D-$\pi$-A materials have attracted a lot attention due to their potential in the creation of flexible and remarkably light solar cells, with low manufacturing cost and high power conversion efficiency (PCE){\cite{hedley2016}}. A key issue to understanding the CT process is the ability to make quantitative predictions and measurements of the characteristics associated to the individual molecular systems that allow useful information for a direct comparison of the electron dynamics inferred in electron and photochemistry, at the nanometric, molecular and electronic scales{\cite{Abendroth}}.

Currently, the increasing availability of kinetic data of CT processes and the development of computational tools allow the study of different molecular systems independently of their size, that exhibit better photophysical properties, and facilitate a direct comparison between theory and experiment{\cite{Ferretti, Cacelli, Prytkova}}. These systems consist of covalent bonds of electron-active chemical species, whether they are electron-donors or electron-acceptors, which can be either connected directly or through a $\pi$-conjugated bridge. The derivatives of the Rhodanine implemented in the synthesis of push-pull systems are an example of these type of systems, and they have been used as an electron-acceptor fragment in a variety of organic compounds of interest; for example, in non-linear second order analytical  reactives, and, more recently, as metal-free organic dyes in the manufacturing of dye-sensitized solar cells (DSSCs){\cite {kuang2008, Zhang2016, Zhu2017}}. For this purpose, the push-pull molecular system of the donor-rhodanine type is efficiently anchored to the meso-porous surface of TiO$_{2}$. The light absorbed by the dye injects electrons into the conduction band of the TiO$_{2}$, thus generating an electric current, while the fundamental state of the dye is regenerated by the electrolyte{\cite{liang2007}}.

The precise prediction of the electron transfer rate in chemical and biological reactions of this type makes them attractive systems for different applications in the field of molecular electronics{\cite{Nar, Page}}. In this work, we present results obtained for new ``push-pull'' chromophores based on derivatives of rhodanine, whereby 4-dimethylamine and 2-formyltetratiafulvalene exhibit the role of electron-donor groups in which the nature of the electron transfer processes is studied when they are connected to a dicyanorhodanine electron-acceptor through a small molecular bridge. To determine the most stable structure, the absorption spectrum and the first electronic state of the complexes were calculated by means of density functional theory (DFT) numerical simulation. We  study the way the electron transfer in complexes gets affected by the presence of a solvent that acts as an environment (also in gas phase), seeking to report on novel quantitative results for such compounds {\cite{Insuasty}}, and explore their potential in the application and design of innovative and highly efficient donor-acceptor multifunctional devices that exhibit optimal electronic properties.  

\section{Chromophores and computational details}
 
Here, we consider the molecules (Z)-2-((1,1-Dicyanomethylene)-5-(4-dimethylamino)benzylidene)-1,3-thiazol-4 (molecule 1), and (Z)-2-((1,1-Dicyanomethylene)-5-(tetrathiafulvalene-2-ylidene)-1,3-thiazol-4 (molecule 2), as shown in Fig. {\ref{fig1}}. The geometries obtained for such most stable conformations (see Supplementary section) were used as input data for the full optimization of calculations of the ground state by means of the hybrid functional B3LYP with a base set 6-31G+, using Gaussian 09 {\cite{frisch2009}}; the corresponding optimized structures are used for the molecules energy calculation. The molecules excited states were calculated by means of the time-dependent density functional theory (TD-DFT), and the results here reported were carried out with the molecules i) in the gas phase, and ii) by simulating an environment--methanol as solvent. In order to see the effects caused by the solvent on the electronic properties of the different compounds, the working molecules in the solvent were designated as follows: system S1: Molecule 1 + Methanol and system S2: Molecule 2 + Methanol, while the system GP1: Molecule 1 in  gas phase and the system GP2: Molecule 2 in gas phase. Additionally, different properties of these molecules, such as higher occupied molecular orbitals (HOMOs), lower unoccupied molecular orbitals (LUMOs), energy gap, reorganization energy, Gibbs free energy, and excitation energy are derived from the computational results. In our theoretical calculations we take into account the effects due to the solvent, since we aim to make occurate predictions that compare with the reported experimental spectra. We consider methanol as solvent with $\epsilon = 32.6$, following a Conductor-like Polarization Continuum Model (C-PCM){\cite{takano2005, chiu2016}}.

\begin{figure}[h]
\centering
\includegraphics[scale=0.6]{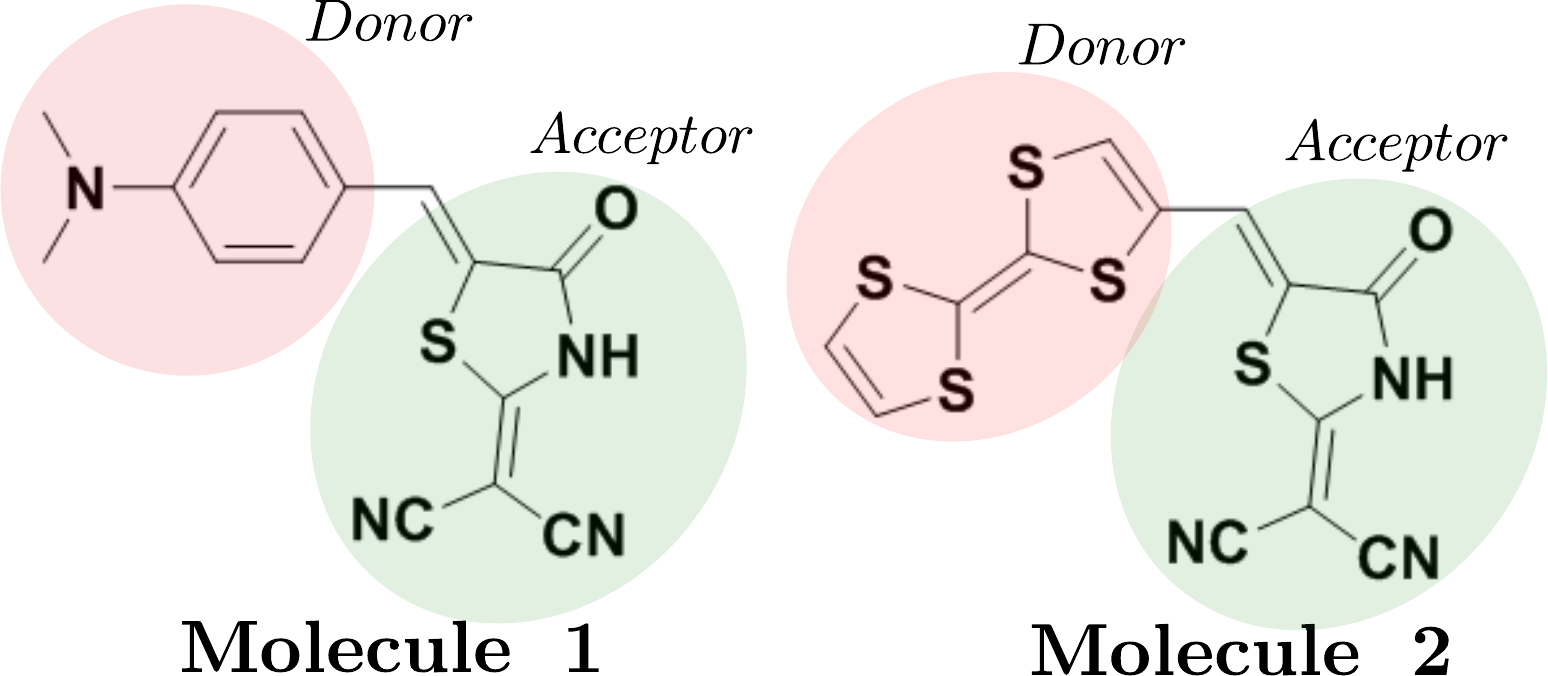} 
\caption{Molecular structure of the chromophores under study. Molecule 1 is the (Z)-2-((1,1-Dicyanomethylene)-5-(4-dimethylamino)benzylidene)-1,3-thiazol-4, and molecule 2 is the (Z)-2-((1,1-Dicyanomethylene)-5-(tetrathiafulvalene-2-ylidene)-1,3-thiazol-4.}
  \label{fig1}
\end{figure} 

\section{Results and discussion}

\textbf{Electronic Transitions}. Computational calculations to optimize the geometry of molecules 1 and 2 were carried out using DFT-B3LYP/6-31G+; subsequently we used TD-DFT for determining the transition states that most favoured the charge transfer and search for a favorable environment in the CT processes. The systems under study are of the donor-acceptor type in which two different electron-donor fragments are connected to an electron-acceptor fragment for generating and effective CT process, which results in a charge separation state for analyzing the key molecular properties in the calculation of the charge distributions in these molecules {\cite{Insuasty}}.

\begin{figure*}[htp]
\begin{center}
\includegraphics[scale=0.51]{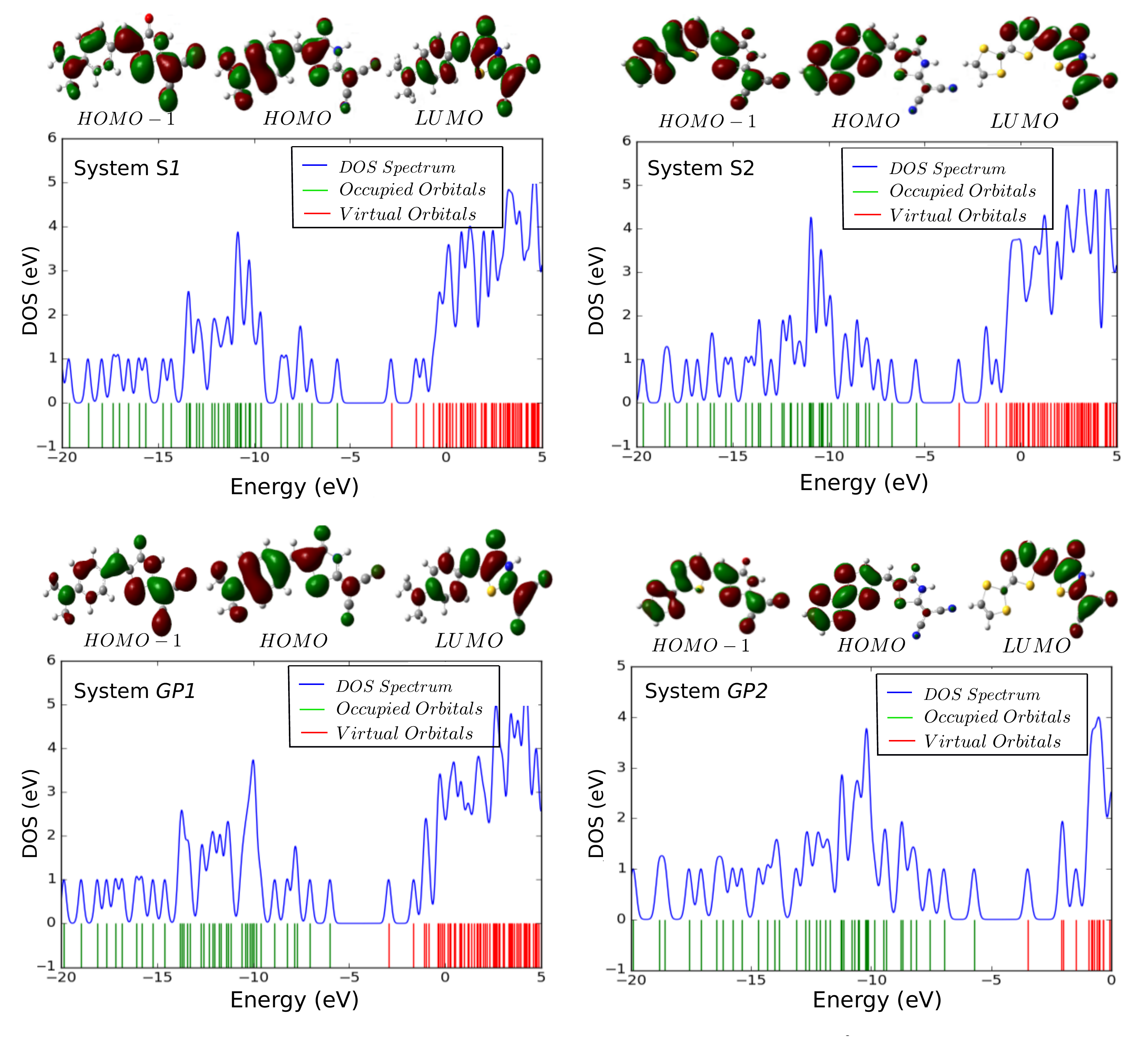}
\caption{Spectra for the total density of states (DOS) obtained using B3LYP/6-31G+ for the systems S1, S2, GP1 and GP2. The blue curve represents the density of state spectrum, the green lines represent the occupied molecular orbitals, and the red lines are the virtual molecular orbitals.} 
\label{fig2}
\end{center}
\end{figure*}

D-$\pi$-A molecular systems have two possible mechanisms for charge transfer, i) super exchange: the charge or electrons transferred do not reside directly in the molecular bridge and the states occupied by the molecule during this time are known as virtual excitations; and ii) sequential charge transfer (hopping): there are real intermediate states that are energetically accessible, and this (thermally activated) mechanism is generally more efficient for long-distance electronic transfer processes{\cite {Ortiz}}.

Here we analyze the frontier molecular orbitals in inder to quantify the relationship between structural and electronic geometry. For system in solvent S1, the HOMO is mostly concentrated in the donor (4-dimethylamino group), while the HOMO-1 and the LUMO are mostly located in the acceptor (2-(1,1-dicyanomethylene)-1,3-thiazole-4), as seen in Fig. {\ref{fig2}}. Therefore, the CT is the charge transfer mixture within the 4-dimethylamino coupled with the CT from 2-(1,1-dicyanomethylene)-1,3-thiazole-4 to 4-dimethylamino moiety. For system S2, the HOMO and HOMO-1 are mainly located in the tetrathiafulvalene moiety, while the LUMO is located in the 2-(1,1-dicyanomethylene)-1,3-thiazole-4 (Fig. {\ref{fig2}}). Thus, the transitions of system S2 from the HOMO to the LUMO together with HOMO-1 to LUMO have a more significant character in the charge transfer with respect to system S1, reflecting that the tetrathiafulvalene acts as a better electron-donor fragment than the first one.

When we observe the frontier molecular orbitals of molecules in gas phase, we see that the behavior described above is maintained. However, their density of state spectra have very significative changes in the spectral densities corresponding to each energy level, which indicates a considerable effect due to the solvent on the electronic and geometric structure of the compounds.

\begin{figure}[htp]
\includegraphics[width=\columnwidth]{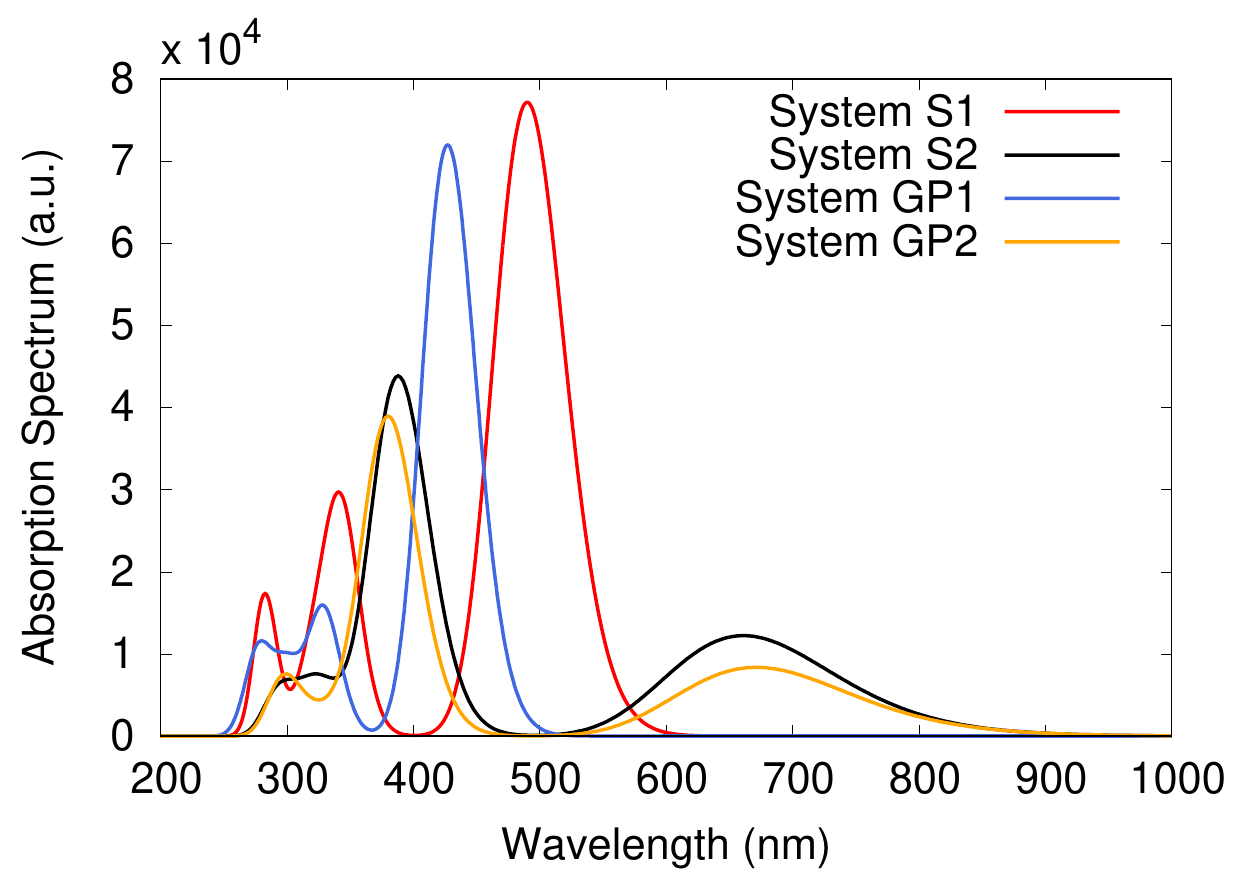}
\caption{Absorption spectrum of the molecules in solvent (systems S1, S2), and molecules in gas phase (systems GP1 and GP2).}
\label{fig3}
\end{figure}

\textbf{Absorption Spectra}. The electronic transition energies and the charge transfer transitions are calculated using TD-DFT/B3LYP \cite{ganji2015,ganji2016}. The UV-Vis absorption spectra for the systems S1, S2, GP1, and GP2, shown in Fig. {\ref{fig3}}, are compared with the experimental spectra reported by Insuasty \cite{Insuasty} in methanol solution to validate these computational results. This comparison shows that the spectral profiles match each other within reasonable accuracy. Here, we observe both experimental and computational spectra with a Full-Width Half Maximum (FWHM) of around $100$ nm.  Additionally, we observe that the experimental spectrum shows a maximum absorption wavelength at $439.00$ nm for S1, and the calculated one is $490.01$ nm.  This difference ($\sim 49$ nm) between the experimental measurements and the calculations takes place by the overestimation of the energy values in the UV/Vis spectrum, and excitation energy of the molecular system by the chosen computational method  \cite{castro2013, Eugenia2013}. Moreover, we show the calculated values of $\lambda_{max}$ corresponding to different solvents. These results confirm the reliability of our method in agreement with the experimental results, see Table T1 (supplementary information).  We find that system S1 has a strong absorption band at $490.01$ nm, together with other bands of lower energy at $343.37$ nm and $283.00$ nm corresponding to transitions of the type $\pi-\pi^{*}$, which are associated to the S$_2$ and S$_7$ states, respectively (see Table {\ref{tab1}}). The intense high energy transition at $490.01$ nm is described by the excitation HOMO$\rightarrow$ LUMO (99\%), according to the orbital transition diagram (See Supporting information S1). This high energy transition can be assigned to intramolecular charge transfer from the 4-dimethylamino electron-donor fragment to the electron-donor fragment; the low energy transition at $343.37$ nm corresponds to the transition HOMO$\rightarrow$ LUMO+1 (65\%). The electronic transitions can be seen as a contribution to the intramolecular charge transfer process from the electron-donor to the electron-acceptor, and like the low energy transition at $283.00$ nm which is described by HOMO$\rightarrow$ LUMO+3 (75\%).
\medskip

\begin{table*}[htp] 
\begin{center}
\centering
\caption{Wavelengths of the most important simulated transition states $\lambda$, oscillator strengths $f_{os}$, excitation energy $E$, and band gap energy $\Delta E_{H-L}$.}
\scalebox{0.8}{
\begin{tabular}{ccccccl} 
\hline  
&&&&&&\\
\multicolumn{1}{m{1.4cm}}{\centering System} & \multicolumn{1}{m{1.0cm}}{\centering States} & \multicolumn{1}{m{1.5cm}}{\centering\textbf{$\lambda$} (nm)} & \multicolumn{1}{m{2.3cm}}{\centering\textbf{$\Delta E_{H-L}$} (eV)} & \multicolumn{1}{m{1.0cm}}{\centering\textbf{$f_{osc}$}} & \multicolumn{1}{m{1.3cm}}{\centering $E$ (eV)} &  Composition \\ 
&&&&&&\\
\hline
             & $S_{1}$     & 490.01 &      & 0.9727 & 2.530 & HOMO$\rightarrow$LUMO (99\%) \\ 
\textbf{S1}  & $S_{2}$     & 343.37 & 2.74 & 0.3391 & 3.611 & H-1$\rightarrow$LUMO (\%34); HOMO $\rightarrow$ L+1 (65\%)  \\ 
             & $S_{7}$     & 283.00 &      & 0.1336 & 4.381 & H-3$\rightarrow$LUMO (16\%);HOMO$\rightarrow$L+3 (75\%) \\ 
\hline
             & $S_{1}$ & 561.10 &      & 0.1995 & 1.875  & HOMO$\rightarrow$LUMO (99\%) \\ 
\textbf{S2}  & $S_{4}$ & 388.31 & 2.27 & 0.6559 & 3.193  & H-1$\rightarrow$LUMO (88\%); HOMO $\rightarrow$L+1 (10\%) \\ 
             & $S_{6}$ & 325.98 &      & 0.1099 & 3.803  & H-2$\rightarrow$LUMO (92\%) \\ 
             & $S_{10}$& 294.66 &      & 0.0962 & 4.208  & H-3$\rightarrow$LUMO (80\%); H-1 $\rightarrow$ L+1 (13\%)\\
\hline 
             & $S_{1}$ & 427.12 &      & 0.9075 & 2.903  & HOMO$\rightarrow$LUMO (99\%) \\
\textbf{GP1} & $S_{2}$ & 329.19 & 3.05 & 0.1939 & 3.767  & H-1$\rightarrow$LUMO (70\%); HOMO $\rightarrow$ L+1 (29\%) \\
             & $S_{7}$ & 281.39 &      & 0.0833 & 4.406  & H-2$\rightarrow$LUMO (76\%) \\
\hline
             & $S_{1}$ & 671.27 &      & 0.1365 & 1.847  & HOMO$\rightarrow$LUMO (99\%) \\
\textbf{GP2} & $S_{2}$  & 380.16 & 2.23 & 0.6322 & 3.261  & H-1$\rightarrow$LUMO (98\%) \\
             & $S_{3}$  & 297.93 &      & 0.1194 & 4.161  & H-3$\rightarrow$LUMO (79\%); H-1 $\rightarrow$ L+1 (11\%) \\ 
\hline 
\end{tabular}}
\label{tab1}
\end{center}
\end{table*}

The system S2 shows a band of high energy absorption at $388.31$ nm and bands of low absorption at $561.10$ nm, $325.98$ nm, and $294.66$ nm. The low energy transition $S_{1}$ for the system S2 is described by the transition HOMO$\rightarrow$LUMO (99\%), which represents an intramolecular CT from the 2-tetrathiafulvalene electron-donor moiety to the 2-(1,1-dicianometilen)-1,3-tiazol-4 electron-acceptor fragment; for the case HOMO-1 and LUMO the latter mostly located in the 2-(1,1-dicianometilen)-1,3-tiazol-4 corresponds to the low transition energy of the state $S_{4}$ described by the transition HOMO-1$\rightarrow$LUMO (88\%) where a CT from electron-donor to electron-acceptor is present. For HOMO-3 and LUMO both mainly concentrated in 2- (1,1-dicyanomethylene)-1,3-thiazole-4, associated with the transition HOMO-3$\rightarrow$LUMO (80\%), a CT within the electron-acceptor and not a transfer from the electron-donor to the concise electron-acceptor takes place.

For the case of the system GP1, comprising the molecule 1 in gas phase, we observe that it presents a high energy absorption band at $427.12$ nm corresponding to the transition state of the HOMO$\rightarrow$LUMO ($99$\%). In addition, as in the case of the system S1, it presents two low energy absorption bands at i) $329.19$ nm, corresponding to the transition state of the HOMO-1$\rightarrow$LUMO ($70$\%) that describes a contribution to the process of CT in the interior of (2-(1,1-dicyanomethylene)-1,3-thiazol-4), since the HOMO-1 and LUMO are more concentrated in the acceptor, with a small CT contribution from the donor to the acceptor (as can be seen in Supporting information {S2}); and ii) a low energy band at $281.39$ nm, corresponding to the transition of the HOMO-2 $\rightarrow$LUMO. By comparing the graphs of systems S1 and GP1, we obtain that the solvent produces an effective shift or bathochromic displacement with respect to the wavelength of absorption.

The system GP2 shows a similar behavior to that of system S2. However, a bathochromic shift is observed in relation to the absorption spectrum when comparing the systems S2 and GP2. The resulting shift is associated to the molecule-solvent interaction, since for the system GP2 this presents a high energy band at $380.16$ nm, and for system S2 in presence of methanol this occurs at $388.31$ nm. In addition, the system GP2 presents two low energy bands, one at $671.27$ nm and the other one at $297.93$ nm; they are associated with the transition states HOMO $\rightarrow$LUMO (99\%), and HOMO-3$\rightarrow $LUMO (79\%), respectively. The energy band at $380.16$ nm is associated to the transition states of HOMO-1 $\rightarrow$LUMO (98\%).

The oscillation strength for an electronic transition is proportional to the transition dipole moment. In general, a large oscillator strength corresponds to large experimental absorption coefficients or a stronger fluorescence intensity {\cite{Sun}}. Table {\ref{tab1}} shows that oscillator stregth ($f_{os}$) values that correspond to the transition states HOMO $\rightarrow$LUMO of systems S1 and GP1, are similar to each other, as are for systems S2 and GP2. The results shown in Table {\ref{tab1}} reveal that the systems S1 and GP1 correspond to systems that have a higher absorption capacity when compared to systems S2 and GP2.

\textbf{Emission Properties}. We use TD-DFT, with the hybrid B3LYP and basis set 6-31G+ in order to compute for the structure in an excited state and simulated the emission spectrum of the systems under study. The maximum emission wavelengths are shown in Table {\ref{tab2}}. The transitions $ S_{1}\rightarrow S_{0}$ and $ S_ {3}\rightarrow S_{0} $ represent fluorescence peaks in the emission spectrum; in addition, the system S1 has the highest oscillator strength, which corresponds to a LUMO$\rightarrow$HOMO transition.

\begin{table*}[htp] 
\begin{center}
\centering
\caption{Emission spectrum results obtained for the systems under study in solvent (S) and gas phase (GP).}
\scalebox{0.88}{
\begin{tabular}{cccccccc} 
\hline  
\multicolumn{1}{m{2cm}}{\centering System} & Electronic Transition &\multicolumn{1}{m{2cm}}{\centering\textbf{$\lambda_{max}$} (nm)} & \multicolumn{1}{m{1.5cm}}{\centering\textbf{$E_{s}$} (eV)} & \multicolumn{1}{m{1.5cm}}{\centering\textbf{$f_{os}$}} & \multicolumn{1}{m{2.6cm}}{\centering Stokes Shift (nm)} &\multicolumn{1}{m{2cm}}{\centering\textbf{$\tau_{R}$} (ns)}  \\ 
\hline  
\textbf{S1}     & $s_{1} \rightarrow s_{0}$ &    508.90      &   2.436    &   1.0126   & 18.0 & 3.85  \\ 
\textbf{S2}     & $s_{3} \rightarrow s_{0}$ &    416.78      &   2.975    &   0.5316   & 28.8 & 5.02  \\  
\textbf{GP1}    & $s_{1} \rightarrow s_{0}$ &    443.28      &   2.797    &   0.9180    & 16.2 & 3.36 \\
\textbf{GP2}    & $s_{3} \rightarrow s_{0}$ &    415.36      &   2.985    &   0.3792   & 35.2 & 7.29  \\ 
\hline
\end{tabular}}
\label{tab2}
\end{center}
\end{table*}

The results for the excitation energy, oscillator strength and radiative lifetime are presented in Table {\ref{tab2}}. We also report the Stokes shift values, defined as the difference between $\lambda_{max}$ of absorption and $\lambda_{max}$ of emission spectrum. The Stokes shift gives the energy difference that exists between the absorption and emission due to the same levels. This provides information about the probability of radiative and non-radiative de-excitation between two levels, where the probability of radiative de-excitation increases with the difference of energy and that of the non-radiative one decreases. Hence, the first one dominates when the energy levels are well separated and the second one does it when we have closer levels. Thus, the radiative lifetime was calculated for the spontaneous emission spectrum using the Einstein transition probabilities according to the expression {\cite{hlel2015, bourass2016}}:

\begin{equation}
\tau_{R} = \frac{c^{3}}{2(E_{fl})^{2}f_{osc}},
\end{equation}

\noindent where $c$ is the speed of light in vacuum, $E_{fl}$ is the fluorescence excitation energy, and $f_{osc}$ is the oscillator strength.

We conclude, as can be seen from Table {\ref{tab2}} that the presence of the solvent favours the radiative processes for the case of molecule 1, since the Stokes shift is greater in the presence of methanol ($18.0$ nm) than in the gas phase ($16.2$ nm). However, for the case of molecule 2 this process is favoured in the gas phase rather than in the presence of the solvent, which is made evident by the longer radiative lifetime found for the gas phase. It is well known that short radiative lifetimes lead to a high efficiency of light emission, while long radiative lifetime facilitates electron and energy transfer. In our case, the radiative lifetime is shorter for systems with higher oscillator strength, which leads to an increase in luminescent efficiency.

The duration of emission ($\tau_{R}$) for the studied molecules have the following order: $\tau_{R}^{GP2}>\tau_{R}^{S2}>\tau_{R}^{S1}>\tau_{R}^{GP1}$. This hierarchy indicates that the change of a donor unit strongly decreases the emission lifetime of the compound in both gas phase and solvent; we find the highest oscillator strength and the smallest lifetime radiation in the case of the systems S1 and GP1, which correspond to the molecule 1 under different environmental conditions. Consequently, molecule 1 represents a good emission material with high efficiency {\cite{zhang2012}}.

\begin{figure}[htp]
\includegraphics[scale=0.4]{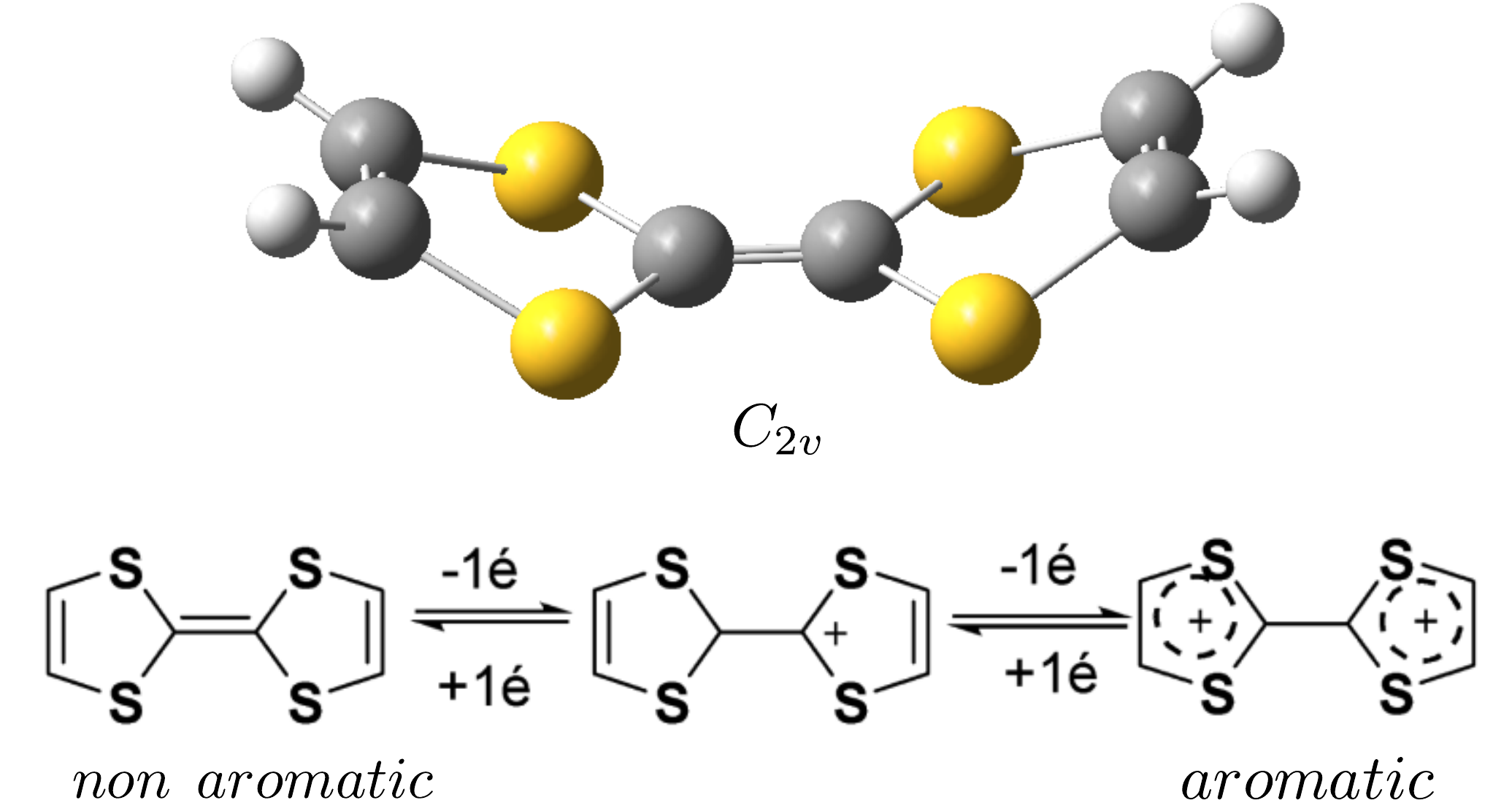}
\caption{Molecular structure of the Tetrathiafulvalene (TTF) neutral unit and the representation as aromatic of the radical cation and dication in the Hückel sense, showing that while TTF$^{+\bullet}$ and TTF$^{2+}$ have a planar D2h symmetry, neutral TTF has a boatlike equilibrium structure with C$_{2v}$ symmetry}
\label{figMM}
\end{figure}

 The previous results show that molecule 2 exhibit higher energy than molecule 1, in agreement with the absorption and emission spectra values for both in the gas phase (GP) and the solvent presence (S). Also, we observed that the radiative times are longer for molecule 2 than for molecule 1.  We attribute it mainly to the fact that in the HOMO-level the molecule 2 shows a charge distribution in the donor fragment while the molecule-1 shows a probability distribution in all around the molecular complex, therefore, the charge transfer process will take more time in the molecule 2 than in 1 as we calculated.  In the molecule 1 case, the electron-donor fragment is the N,N-dimethylbenzene, which has an aromatic character in its ground state, i.e., it complies with the Hückel rule of $4N+2\pi$ electrons in the benzene ring. If the system experiencing photo-excitation, and the phenomenon of intramolecular charge transfer occurs, the aromaticity is compromised and competes with the charge separation process acquiring radical-cation characteristics in such a process. These dynamics increases the radiative process in the vacuum and within a solvent avoiding the structural variation counteract this phenomenon.  In contrast, the molecule 2, which has the tetrathiafulvalene unit (TTF) as the electron-donor fragment, this unit is a non-aromatic fragment of 14 $\pi$ electrons and structural symmetry $C_{2V}$ that deviates from the planarity as shown in Fig. \ref{figMM} \cite{segura2001}.  In the photoexcitation process, this donor-fragment is capable of sequentially providing electrons sequentially until reaching the stable state which implies a dicationic state in a planar structure with aromatic behavior. This behavior decreases the radiative processes and improves the interaction with the solvent generating charge separated species with greater stability.

\textbf{Charge Transfer Rate}. Charge transfer is a crucial process involved in many physical and biological phenomena such as the photosynthesis {\cite{Gray}}; this process can be estimated, in a first approximation, by using the semi-classical theory of Marcus {\cite{Marcus1, Marcus2}}:

\begin{equation}
k_{e(h)} = \frac{2\pi}{\hbar} \frac{\vert V_{e(h)}\vert^{2}}{\sqrt{4\pi\lambda_{e(h)} k_{B}T}} \exp\left(-\frac{ \lambda_{e(h)}}{4 k_{B}T} \right),
\label{ecu1}
\end{equation}
 
\noindent where $V_{e(h)}$ is the electronic coupling between the final and the initial state for electrons (holes), $\lambda_{e(h)}$ is the reorganization energy for electron (hole), and $k_{B}$ denotes Boltzmann constant.
 
For an efficient CT mechanism the reorganization energy of the molecular system must be small and an electronic coupling between the electron-acceptor and electron-donor parts is necessary {\cite{Coropceanu2007}}. The reorganization energy comprises two factors, the first one is the internal or intramolecular reorganization energy ($\lambda_{int}$), and the second one is the external or intermolecular reorganization energy ($\lambda_{ext}$). The $\lambda_{int}$ accounts for the structural changes between neutral and ionic states and must be calculated, while $ \lambda_{ext}$ reflects the change in the polarization of the medium after the CT takes place {\cite{li2015}}. 

The intramolecular reorganization energy $\lambda_{int}$ can be estimated for the electrons (reorganization energy of electron $\lambda_{e}$) and for the holes (reorganization energy of holes $\lambda_{h}$), and can be expressed by the following equation {\cite{pandith2014, Kose, DeVine}}:
\vspace{-1.2cm}

\begin{eqnarray}
&\lambda_{e}& = \lambda_{1}^{e}+ \lambda_{2}^{e} = \left( E_{0}^{-} - E_{-}^{-} \right) + \left( E_{-}^{0} - E_{0}^{0} \right)\nonumber\\
&\lambda_{h}& = \lambda_{1}^{h}+ \lambda_{2}^{h} = \left( E_{0}^{+} - E_{+}^{+} \right) + \left( E_{+}^{0} - E_{0}^{0} \right),
\label{ecu2}
\end{eqnarray}

\noindent where $ E_{0}^{+}$($E_{0}^{-}$) is the energy of the cation (anion) calculated with the optimized structure of the neutral molecule. Similarly, $E_{+}^{+}$($E_{-}^{-}$) is the energy of cation (anion) calculated with the optimized cation (anion) structure, $E_{+}^{0}$ ($E_{-}^{0}$) is the energy of the neutral molecule calculated at the cationic (anionic) state. Finally, $E_{0}^{0}$ is the energy of the neutral molecule at the ground state.

\begin{table*}[htp] 
\begin{center}
\centering
\caption{Molecular calculation of the reorganization energy for electrons (holes) $\lambda_{e(h)}$, electronic coupling for the electron (holes) transfer mechanism $V_{e (h)}$, and electron (hole) transfer rate $k_{e(h)}$.}
\medskip
\scalebox{0.9}{
\begin{tabular}{ccccccc}
\hline 
\multicolumn{1}{m{1.5cm}}{\centering System} & \multicolumn{1}{m{1.5cm}}{\centering \textbf{$\lambda_{e} $}(eV)} & \multicolumn{1}{m{1.5cm}}{\centering \textbf{$\lambda_{h} $}(eV)} & \multicolumn{1}{m{1.5cm}}{\centering \textbf{$V_{e} $}(eV)} & \multicolumn{1}{m{1.5cm}}{\centering \textbf{$V_{h} $}(eV)} &  \multicolumn{1}{m{2.5cm}}{\centering \textbf{$k_{e}(10^{15}$} s$^{-1}$)}            & \multicolumn{1}{m{2.5cm}}{\centering $k_{h}(10^{15}$ s$^{-1}$)} \\ 
\hline 
S1  & 0.184        & 0.094        & 0.660   & 0.685  & $2.87$ & $5.13$  \\ 
S2  & 0.340        & 0.219        & 0.700   & 0.620  & $0.53$ & $1.67$  \\
GP1 & 0.271        & 0.151        & 0.655   & 0.520  & $1.01$ & $2.72$  \\ 
GP2 & 0.374        & 0.221        & 0.720   & 0.630  & $0.37$ & $1.61$  \\
\hline 
\end{tabular}}
\label{tab3}
\end{center}
\end{table*}

\begin{figure}[htp]
\centering
\includegraphics[scale=0.5]{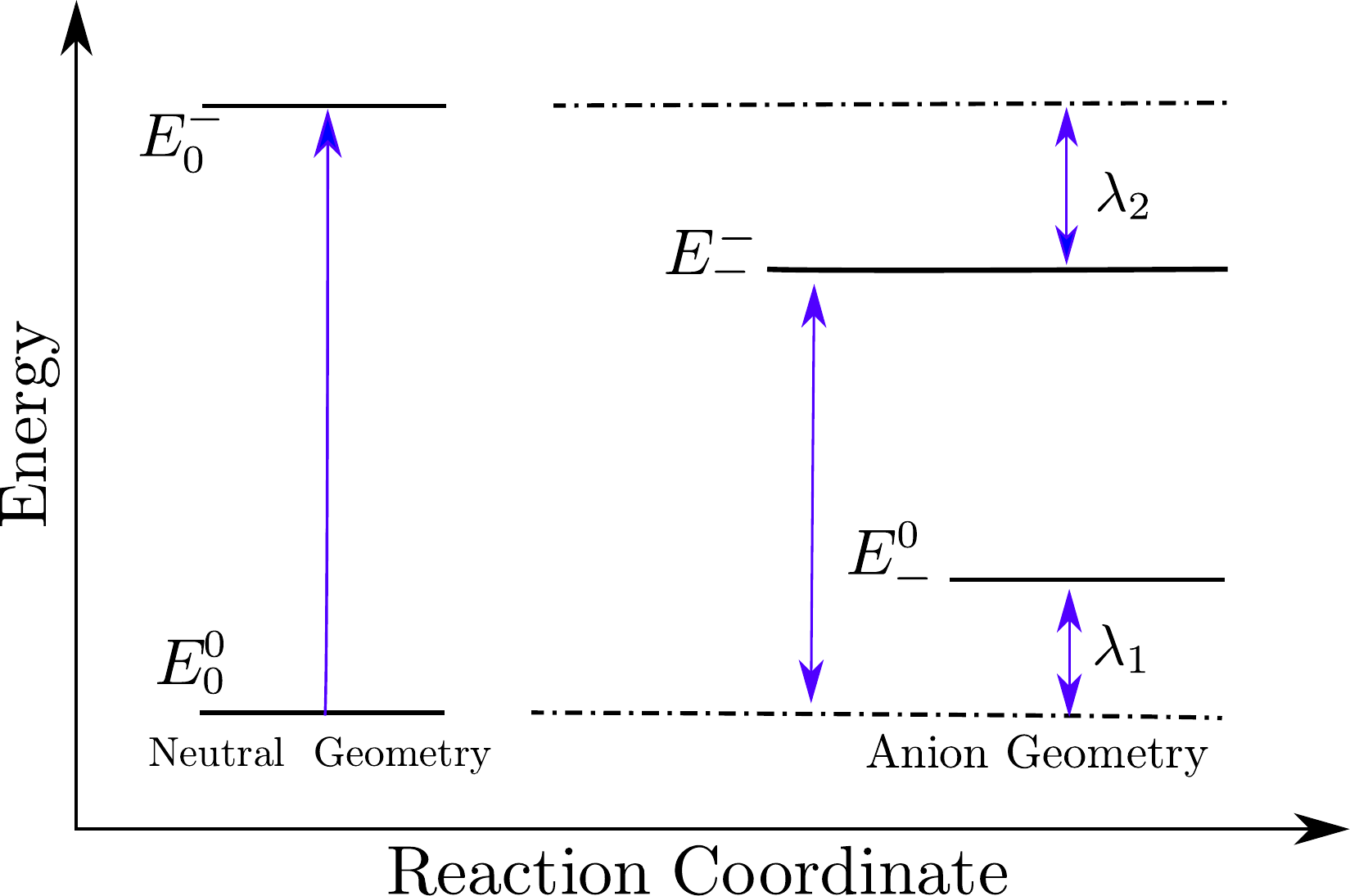}
\caption{Scheme for the calculation of the reorganization energy for the electron transfer; $ \lambda_{1} $ is the reorganization energy of the neutral molecule and $ \lambda_{2}$ denotes the reorganization energy of the radical anion.}
\label{fig6}
\end{figure}

This redistribution of energies for the case of electron reorganization energy is best observed in Fig. {\ref{fig6}}. The external reorganization energy $ \lambda_{ext} $ explains the nuclear shifts in the surrounding medium and the resulting electronic effects are much harder to calculate. This is assumed to be, for many authors, between $ 0.2 $ eV  and $ 0.5 $ eV for simple models based on the dielectric properties of organic matrices {\cite{Olivier, Kose}}.  

We calculate the electron reorganization energy using Fig. {\ref{fig6}}. According to Marcus model, the rate of electron transfer depends mainly on the energy of reorganization and the coupling between the donor and the acceptor, in addition to the general exergonity of the process \cite{pandith2014}. It is also considered that for a self-exchange electron transfer process (where interaction with solvent is not considered) the change in the Gibbs free energy is zero and the electron transfer rate will only depend intrinsically on the barrier of activation, marked by the internal and external reorganization energy, and the electronic coupling parameter $V$. The $ \lambda_{int} $ in Eq. {(\ref{ecu2})} for the self-exchange process has two contributions, arising from the geometric relaxation along inter-nuclear coordinate upon moving from neutral-state to the charged-state geometry and vice versa. The electronic coupling $V$, another key parameter in the CT process, is the geometrically most dependent element of the kinetic constant because its value depends on the distance between donor-acceptor and the geometry of the system (orientation of the orbitals); this parameter is also sensitive to changes in the systems under study such as solvents, and temperature, among others {\cite{Kani, Li3, Valeev, Mallick}}.

Here, we use the generalized Mullinken-Hush method (GMH) {\cite{Cave}} to calculate the electronic couplings, and the operator used in the GMH method is the adiabatic dipole moment matrix $\mu_{12}$ {\cite{Cave}}. Under this approach (and in the weak coupling regime), the electronic coupling for a direct donor-acceptor coupling is calculated as {\cite{Li}}:

\begin{equation}
V=\frac{\Delta E_{12}\mu_{12}}{\sqrt{(\Delta\mu_{1}-\Delta\mu_{2})^{2} + 4\mu_{12}^{2}}},
\end{equation}

\noindent where $ \Delta E_{12}$ is the orbital energy difference, and $\mu_{12}$ is the dipole moments difference of the adiabatic states.

The resuls for the reorganization energy, electronic coupling, and the electron transfer rate are shown in Table {\ref{tab3}}. As reported in {\cite{Ran, Sun, Zou}}, it has been found that at low values of reorganization energy, the transfer rate is high. The hole reorganization energy calculated for the systems S2 and GP2 are smaller than those for the systems S1 and GP1; this implies that the hole transfer rate is greater in the systems S2 and GP2, and we also note that for the case of the system S2 compared to the GP2 the hole transport rate is higher in the presence of methanol than in the case of the gas phase, confirming this behavior for the case of the system S1 and GP1 where the same situation is observed, which indicates that methanol does not favour the transport of holes. Furthermore, the hole reorganization energies $\lambda_{h}$ for all systems are smaller than that of N,N'-diphenyl-N,N'-bis(3-methlphenyl)-(1,10-biphenyl)-4,4'-diamine (TPD), which is a typical hole transport material with $\lambda_{h}=0.290$ eV {\cite{gruhn20021}}.  This implies that the hole transfer rates of the molecules 1 and 2, in the condition under study, might be higher than that of TPD. Thus, the molecules 1 and 2 might comprise good hole transport materials from the stand point of the smaller reorganization energy. On the other hand, we observe that for the case of the system S2 compared to GP2 the hole transport rate is higher in the presence of methanol than in the case of the gas phase, confirming this behavior for the case of the systems S1 and GP1 where the same situation occurs, which indicates that methanol favours the transport of voids.  

The value of $ \lambda_{e} $ is smaller for the case of the systems S1 and GP1 than for the systems S2 and GP2: this indicates that the electron transfer rates for S1 and GP1 will be larger than those due to the systems S2 and GP2 as can be seen in Table {\ref{tab3}}. In addition, by comparing $\lambda_{e}$ for the systems S1 and GP1, as well as for S2 and GP2, we see that the following holds for $\kappa_{e}$: $\kappa_{e}^{S1}>\kappa_{e}^{GP1}$, and $ \kappa_{e}^{S2}> \kappa_{e}^{GP2} $, which indicates that the electron transfer process is favoured by the presence of solvent in the molecules 1 and 2. In addition, by comparing the reorganization energies for electron reorganization and holes, we observe that the values of $\lambda_{h}$ are smaller than those for $\lambda_{e}$, suggesting that the carrier mobility of the electrons is larger than that of the holes. Hence, the molecules 1 y 2 can be used as promising hole transport materials in, e.g., organic light-emitting diodes from the stand point of the smaller reorganization energy, which can be corroborated with the $\kappa$ values shown in the Table {\ref{tab3}}. Finally, given that $\kappa_{e}$ and $ \kappa_{h}$ are greater for the systems S1 and GP1, we conclude that the molecule 1 is better at transporting charge than the molecule 2. In addition to the previous analysis, it can be seen that the molecule 1 is more transport-efficient than the molecule 2.

\subsection{Conclusions}

We have theoretically investigated different optical and electronic properties for new structures based on 4-dimethylamino and tetrathiafulvalene as electron-donor groups and dicyanorhodanine as electron-aceptor group, in which the correlation between structures and electronic dynamics is studied by means of theoretical chemical calculations. It was observed that the presence of a solvent with which the molecules 1 and 2 interacts, favours the electronic transfer process. In addition, the change in the donor group shows that 4-dimethylamino acts as a better electron donor and hole transporter than tetrathiafulvalene, which can be seen by observing their rates of transfer, for both electrons and hole. Also, we observe that if compared the transfer rate of the molecule 1, it is found that this compound is a better promisor for the transport of holes than electrons ($\kappa_{h}> \kappa_{e} $) in both cases, in presence of solvent and gas phase. As regards the wavelength of absorption, this shows a bathochromic effect between the gas phase and the presence of the solvent. The computational results predict the electronic properties of the systems S1, S2, GP1 and GP2, and the analysis of the molecular frontier orbitals shows that the vertical electronic transitions of absorption of the studied compounds are characterized as intramolecular charge transfer. In addition, the molecules 1 and 2 can be used as void transport materials.

\begin{acknowledgement}

The authors acknowledge support by the Colombian Science, Technology and Innovation Fund-General Royalties System (Fondo CTeI-Sistema General de Regalías) under contract BPIN 2013000100007.

\end{acknowledgement}

\begin{suppinfo}

Contains the optimal geometries, the total energy and the frontier molecular orbitals involved in the electronic transitions at the TD-DFT B3LYP/6-31G+ levels.
\end{suppinfo}

\bibliographystyle{acs}
\bibliography{Bibliografia}
\end{document}